\RequirePackage[2020-02-02]{latexrelease}
\documentclass[10pt,pre,twocolumn,amsmath]{revtex4}
\pdfoutput=1
 
\usepackage[pdftex]{graphicx} 
\usepackage{color}

\begin{document}



\title{Electron drag force in EUV induced pulsed hydrogen plasmas}

\author{M. Chaudhuri$^*$, A. Yakunin, M. van de Kerkhof and R. Snijdewind}
\affiliation{ASML The Netherlands B.V., P.O.Box 324, 5500AH Veldhoven, The Netherlands}

\begin{abstract}
Extreme ultraviolet (EUV) induced pulsed plasma is unique due to its transient characteristics: the plasma switches between non-thermal state (when EUV power is ON at the beginning of the pulse) and thermal state (end of the pulse at $\sim$ 20 $\mu$s). It is shown that although electron drag force acting on nm size particles in hydrogen plasma is negligible compared to the ion drag force at the beginning of the pulse, however it can be dominant at the end of the pulse and can play important role in particle transport leading to defectivity issues for semiconductor chip production technologies.  
\end{abstract}

\maketitle


The impact of semiconductor integrated circuits (IC)/chips on modern lifestyle is hard to overstate. From computers to communication, education to entertainments, the growth of electronics technology accompanied by the advancement of semiconductor chip manufacturing technology has been phenomenal. The fascinating evolution from single transistors to current memory chips and microprocessors with billion-transistors is a remarkable story. The fabrication of such ICs involves a great variety of physical and chemical processes performed on a silicon substrate. Fundamental of all of these processes is optical projection lithography which is basically a photographic process where an image of the mask (pattern) is exposed and projected onto the wafer which is covered by a photoresist (light-sensitive polymer). After development, the 3D images of the pattern can be obtained on the substrate~\cite{Mack_litho}. The lenses/optical elements used in the projection optics are almost perfect so that the resulting image is almost aberration free and mainly diffraction-limited. For such images, the resolution is determined by the wavelength of the imaging light ($\lambda$) and the numerical aperture (NA) of the projection lens according to the Rayleigh resolution criterion: $\propto \lambda/$NA. 
To obtain best possible resolution, the EUV lithography has been introduced in recent times which uses highly energetic EUV photons (energy $\sim$ 92 eV) with much shorter illumination wavelength (13.5 nm). One of the unavoidable side effect of this development is the generation of EUV photon induced plasma due to the interaction of such highly energetic photons with the low pressure (1-10 Pa) background hydrogen gas~\cite{BanineJAP2006, van_der_Horst_2014, Roderik_2018, Kerkhof_2019, Becker_2019, Kerkhof_2020, Kerkhof_2021}. It is to be noted that the EUV photon induced plasmas differ from traditional low-temperature laboratory plasmas (DC discharge, capacitively coupled radio frequency plasmas, inductively coupled plasma, dielectric barrier discharges, etc), in a sense that there exists no continuous external power supply to sustain the discharge. Although the photon induced plasmas is an active subject of investigation for astrophysical plasmas for some time~\cite{Goertz1989,Rosenberg1995,Fortov1998,PhysRevLett.84.6034,2001_Weigartner}, the experimental research has gained momentum recently due to industrial applications in the form of EUV lithography.

The spatial and temporal evolution of the EUV plasmas has been investigated experimentally and with particle-in-cell (PIC) model~\cite{van_der_Horst_2016, Astakhov_2016,Beckers_2016,van_de_ven_2018}. When the plasma decay time is much shorter than the EUV pulse repetition time (20 $\mu$s) then the plasma evolution occurs in different stages: at first the plasma creation happens due to direct photoionization of the hydrogen gas (ionization energy of 13.6 eV) during the passage of highly energetic EUV photons. In this phase, the plasma contains highly energetic electrons with excess photon energy and non-Maxwellian energy distribution. A part of these electrons then move towards the wall leaving behind a positive space charge region which confines the remaining electrons and allow positive ions to accelerate towards walls. The confined electrons lose their energy within few tens of nanoseconds due to electron impact ionization and increase the plasma density. After this stage, plasma starts to expand and the local electron density decreases rapidly. During the last stage of the process, the electrons are supposed to reach their equilibrium temperature with plasma density continues to decrease due to ambipolar diffusion and recombination processes at wall. The energy is supplied within few tens of nanoseconds by the EUV pulse while it takes $> $20 $\mu$s for the plasma to completely extinguish which makes the highly transient plasmas between nonthermal ($T_e >> T_i$) and thermal states ($T_e \sim T_i$). This effect leads to a buildup process over multiple pulses. Here, $T_{e(i)}$ are the electron (ion) temperatures. Such transient plasmas generates an electric field ($E$) when it comes in contact with the surface and is also responsible for charging processes of the nano particles which are present in the system as part of the contamination due to lack of sufficient cleanliness of the surfaces or due to other mechanisms such as blistering or spitting. In presence of such transient electric fields, charge dependent volume forces such as electrostatic force ($F_{\rm el}$), ion drag ($F_{\rm id}$) and electron drag ($F_{\rm ed}$) forces play important roles for nano particle dynamics. The goal of this work is to show that the electron drag force which was always neglected earlier in EUV photon induced plasmas can play important roles for nano particle transport towards critical zones under typical operational conditions in EUV lithography machines.

\begin{figure}
\includegraphics[width=0.95\linewidth]{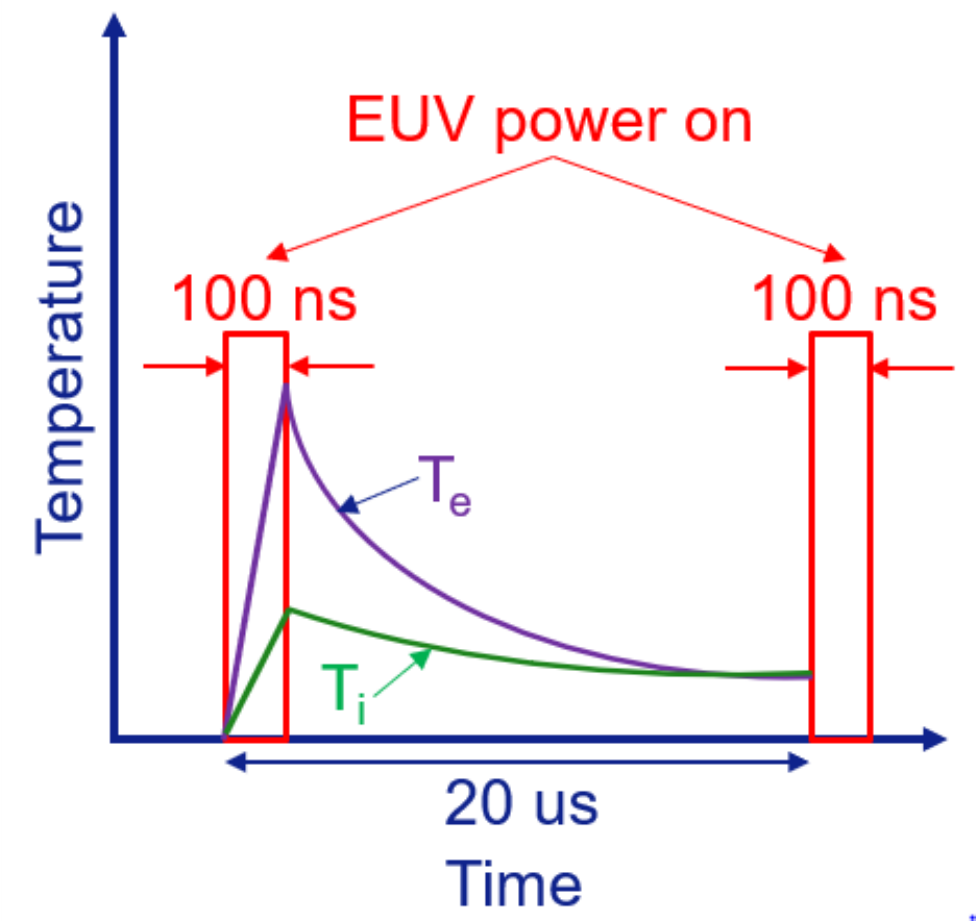}
\caption{A simple illustration of EUV plasma generation using pulsed power mode. Each pulse is 20$\mu$s duration within which the EUV power is ON for 100 ns. During this initial phase of power ON mode, the plasma is non-thermal ($T_e \gg T_i$). After this initial phase, the electron temperature decreases rapidly and at the end of the pulse, the plasma becomes thermal ($T_e \sim T_i$). For each pulse, the plasma exhibits the transient characteristics by switching between non-thermal and thermal states.}
\label{euv_pulse}  
\end{figure}

The electrostatic force $F_{el} = -ZeE$ on the charged particle is generated as direct impact of electric field where $Z$ is the charge number for each individual particle. The indirect effect of electric field is to produce $F_{\rm id}$ and $F_{\rm ed}$ on the grains which are nothing but the momentum transfer rates from drifting ions and electrons to the grains within the electric field. Depending on the plasma regime, different forces respond differently and the competition amongst them are responsible for the equilibrium states and transport properties of the grains~\cite{Fortov2005PR}. Often the ion drag force dominates over the electron drag force because of large ion-to-electron mass ratio and plays very important role to understand particle dynamics in laboratory and microgravity experiments~\cite{PhysRevE.70.056405.khrapak}. However, a complete self-consistent model for ion drag force is yet to be constructed which can describe all cases of interests. Below, we first briefly describe present level of theoretical understanding of the ion drag force and then we justify to use suitable analytical expressions for estimations depending on our experimental parameter regime. There are mainly two approaches to calculate ion drag force: binary collision (BC) approach and linear response (LR) formalism ~\cite{Fortov2005PR}. Each of them has pros and cons, but they are complementary to each other ~\cite{khrapak.pop.2005}. In the BC approach, the velocity dependent momentum transfer cross-section $\sigma(v)$ is obtained after analyzing ballistic ion trajectories in the isotropic attractive Debye-H{\"u}ckel (Yukawa) potential of the particle. Then the ion drag force can be calculated by integrating $\sigma(v)$ with suitable ion velocity distribution function~\cite{Khrapak_PhysRevE.66.046414}. The scattering parameter ($\beta$) is an important quantity which determines the strength of ion-particle coupling for momentum transfer process and is defined as, $\beta = Ze^2/(m_iv^2\lambda)$, where $\lambda$ and $m_i$ denote the effective screening length and ion mass. The advantage for BC approach is that it can be applied for any strength of ion-particle coupling but it completely neglects collisional effects (for example ion-neutral charge exchange collisions) and ion flow effects which creates potential anisotropy. In the LR approach, the ion drag force is calculated by solving Poisson equation coupled to the kinetic (or hydrodynamic) equations for the ions and electrons~\cite{PhysRevLett.92.205007.ivlev, PhysRevE.71.016405.ivlev,chaudhuri.pop.2007,chaudhuri.pop.2008,khrapak.IEEE.2009}. The advantage for the LR approach is that the collisional effects and the ion flow induced potential anisotropy are taken into account self-consistently for ion drag force calculation, but it is applicable only for weak ion-particle coupling $\beta < 1$. However, the dominance of the ion drag force over the electron drag force may change when the electrons drift much
faster than the ions because of their much higher mobility. It was
shown in an earlier work by Khrapak and
Morfill~\cite{Khrapak_PhysRevE.69.066411} that in the collisionless regime the
electron drag force calculated using BC approach can indeed dominate over the electric and ion
drag forces provided the electron-to-ion temperature ratio is not
too high. In this case, using the Coulomb scattering theory for point-like particles with equal magnitude of Coulomb logarithm for electron-particle and ion-particle interactions, the ratio of ion drag and electron drag forces for subthermal drifts can be written as~\cite{Khrapak_PhysRevE.69.066411},
\begin{equation}
\label{id_ed_ratio_quasineutral} 
\frac{F_{id}}{F_{ed}} \sim \gamma \mu^{1/2} \tau^{3/2} 
\end{equation}

\begin{figure}
\includegraphics[width=\linewidth]{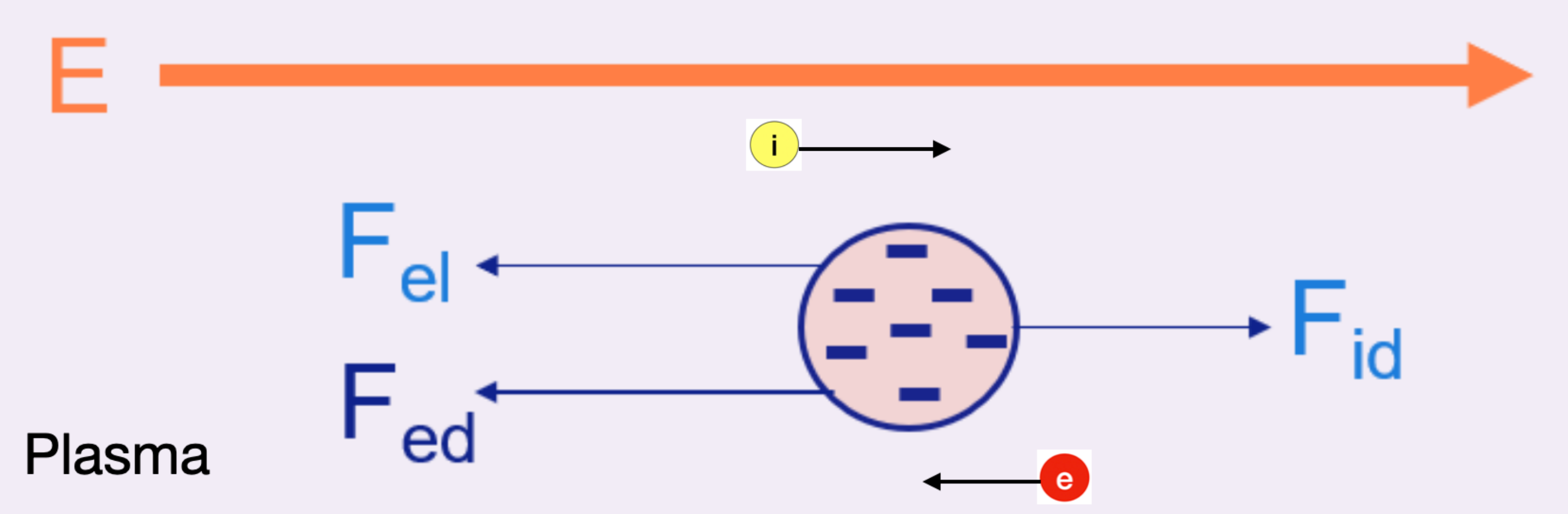}
\caption{An illustration of the charge induced volume forces acting on the negatively charged stationary particle in the presence of an external electric field (E): electric force ($F_{el}$), ion drag force ($F_{id}$) and electron drag force ($F_{ed}$)}. 
\label{force_balance_illustration}  
\end{figure}

Here, $\gamma = u_i/u_e$ is the ratio of ion-to-electron drift velocities, $\mu = m_i/m_e$ is the ion-to-electron mass ratio and $\tau = T_e/T_i$ is the electron-to-ion temperature ratio. For the particle movements in a stationary hydrogen plasma, $u_i \sim u_e$. Then at the beginning of the pulse for non thermal plasma condition, $F_{id} >>F_{ed}$. But then ion drag force decreases with time and $F_{id} \sim F_{ed}$ when $\tau \sim 12$. However, in case of weakly ionized plasmas in an external large scale electric field with mobility limited electron and ion drifts, the electrons drift much faster than ions, $u_e \gg u_i$. Such a situation was considered earlier in the regime of collisionless ion and electron trajectories in the vicinity of the grain (BC approach) and the ratio of ion drag and electron drag force was estimated as~\cite{Khrapak_PhysRevE.69.066411},
\begin{equation}
\label{id_ed_ratio_efield} 
\frac{F_{id}}{F_{ed}} \sim \frac{\omega[1+(z\tau/2)+0.5(z\tau)^2 \ln(2\xi/z\tau)]}{(1+z/2)e^{-z} + 0.5z^2 \ln(2\xi/z)}
\end{equation}
Here, $\omega = \sigma_{en}/\sigma_{in}  $ is the ratio of electron-neutral and ion-neutral collision cross-section. The particle size is normalized by the effective debye length, $\xi = \lambda_{\rm D}/a$. 

Later, Chaudhuri et. al. extended the work of electron drag force to the collisional regime using LR formalism with hydrodynamic approach~\cite{chaudhuri.pop.2007.054503}. In case of weakly collisional ($\ell_i > \lambda_{\rm D}$) and highly collisional plasma ($ \ell_i,  \ell_e < \lambda_D$), ion and electron drag forces (LR) experienced by a nonabsorbing particle are directed along the drift velocities of ions and electrons respectively i.e. they act in opposite directions. The ratio of absolute magnitude of ion-to-electron drag forces acting on a nonabsorbing particle is $|F_{id}/F_{ed}| \sim \tau^2$ which implies that for thermal plasmas, they exactly cancel each other~\cite{chaudhuri.pop.2007.054503}. But for nonthermal plasma, the ion drag force (directed opposite to the electric force) always dominates. However, for absorbing particle in the highly collisional regime, it is possible to obtain negative ion drag force as ion absorption reduces the magnitude of the ion drag force. On the other side, the electron absorption increases the magnitude of the electron drag force. The absolute ratio of ion-to-electron drag force for an absorbing particle is $|F_{id}/F_{ed}| \sim \tau$~\cite{chaudhuri.pop.2007.054503}.

\begin{figure}
\includegraphics[width=\linewidth]{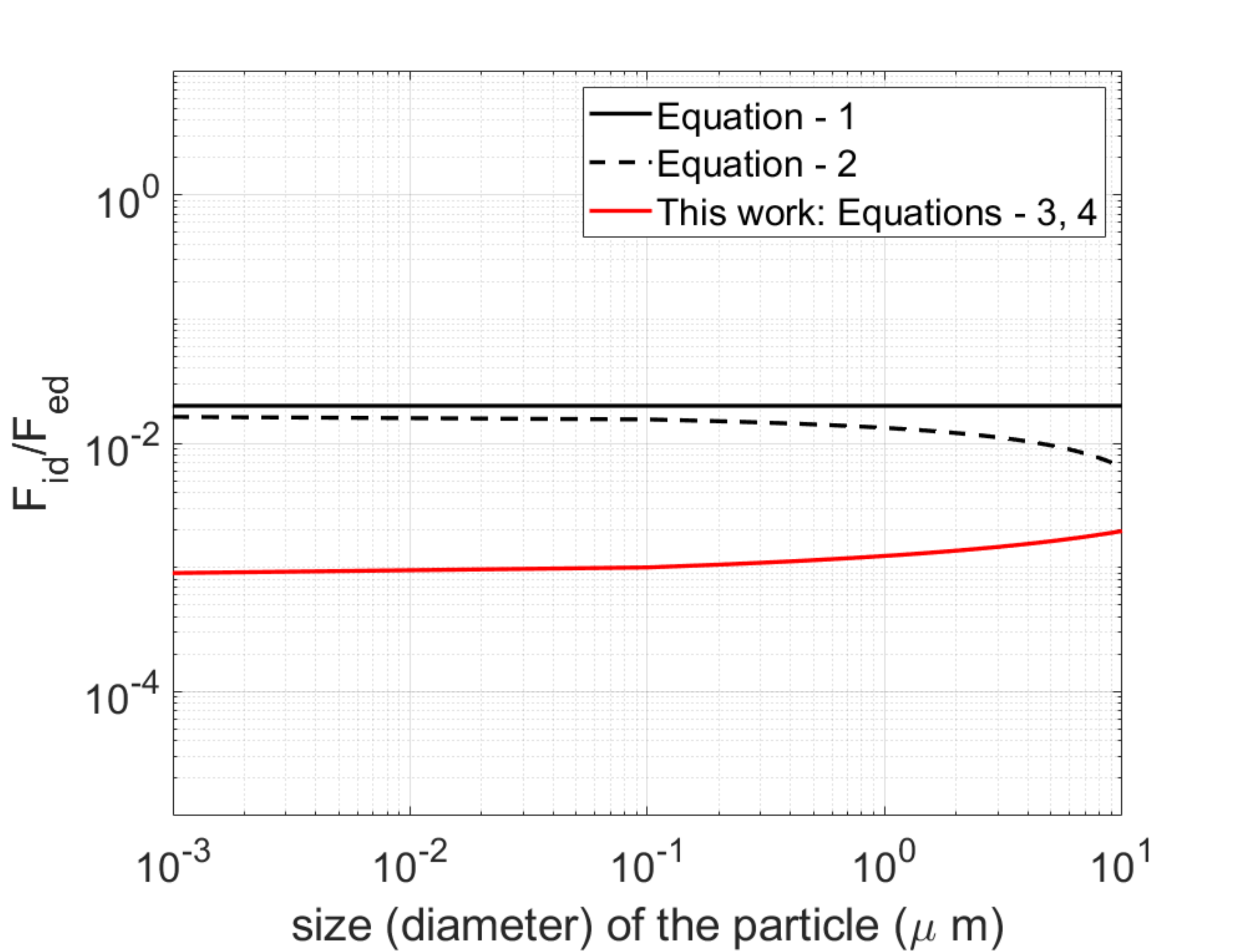}
\caption{The variations of ratio of ion drag force and electron drag force with particle size (diameter) in thermal plasma for three different cases: Eqn-1 (collisionless, subthermal drifts for ions and electrons), Eqn-2 (collisionless, mobility limited drifts with $u_e \gg u_i$)  and this work (collisionless electrons, weakly collisional ions with suprathermal drifts and weak ion-grain coupling). The relevant parameter values have been described in the text.}
\label{force_ratio}  
\end{figure}


To understand the importance of electron drag force for particle transport in transient EUV induced pulsed plasmas, we consider a single pulse duration where the plasma switches between nonthermal (initial stage of the pulse when EUV power is ON with $\tau >> 1$) to thermal states (end of the pulse when $\tau \sim 1$) where $\tau = T_e/T_i$ is the electron-to-ion temperature ratio. 
Typically, the EUV lithography machines operate with following conditions~\cite{Kerkhof_2021}: background hydrogen gas pressure at $p \sim 5$ pa, plasma (electron) density with $n_e \sim 4e8/cm^3$, E $\sim$ 5 V/cm, $ T_e \sim 10 eV$ and  $ T_i \sim 0.025 eV$ (room temperature). For the thermal state, $ T_e \sim T_i \sim 0.025$ eV. 
The charge of a particle is an important parameter which plays an important role on particle dynamics. The orbital motion limited (OML) theory is commonly used to estimate particle charge which assumes collisionless, ballistic ion and electron trajectories in the vicinity of an isolated particle without the presence of any kind of interaction potential barrier~\cite{Allen_1992,kennedy_allen_2003}. In this case, the electron flux can be written as, $I_e \approx \sqrt{8\pi}a^2 n_e v_{Te}\exp(-z)$ where $n_e$ is the electron density, $v_{Te} = \sqrt{T_e/m_e}$ is the electron thermal velocity. The ion flux can be written as, $I_i \approx \sqrt{8\pi}a^2n_iv_{\rm Ti}(1 + z\tau)$ where $a$ is the particle radius, $n_i$ is the ion density, $v_{\rm Ti} = \sqrt{T_i/m_i}$ is the ion thermal velocity and $z = Ze^2/aT_e$ is the normalized particle charge. The estimated charge for a 100 nm diameter particle in nonthermal plasma condition mentioned above is $Z \sim 325$. However, when the experiments are performed at 5 pa background gas pressure then the collisionality index (normalized ion mean free path w.r.t effective debye length), $\eta = \lambda_{\rm D}/\ell_i \sim 0.07$ which represents weakly collisional regime with effective debye length $\lambda_{\rm D} \sim 61 \mu$m. In this regime, the OML expression for the ion flux should not be applied for particle charge determination but rather the collision enhanced ion model should be used, $I_i \approx \sqrt{8\pi}a^2n_iv_{\rm Ti}(1 + z\tau + 0.1z^2\tau^2\eta)$ and the estimated charge for 100 nm diameter particle is $Z \sim 190$. 
The result shows that the charge values in weakly collisional regime are significantly less than that calculated in collisionless regime using OML theory and consistent with previous experimental results~\cite{PhysRevLett.93.085001.Ratynskaia, PhysRevE.72.016406.Khrapak,Khrapak_2012,Antonova_2019}. To estimate the magnitude of electron and ion drag forces, we consider collisionless electron drag force~\cite{Khrapak_PhysRevE.69.066411}, 
\begin{equation}
F_{ed} = \frac{2}{3}\sqrt\frac{2}{\pi}\left(\frac{T_i}{e}\right)^2 \xi^{-2}\tau^{2} (1+\tau)^{-1} M_{e} \Phi(z, \xi)
\end{equation}
Here, $M_{e} = u_e/v_{Te}$ is the thermal mach number for electrons. The parameter $\Phi(z, \xi)$ accounts for the electrostatic interaction between electrons and charged particles which can be estimated as contribution from direct collisions $\approx (1+z/2)e^{-z}$ and from scattering $\approx (z^2/2) \ln(2\xi/z)$. However, for the ion drag force, the expression for weakly collisional regime as was obtained by LR formalism has been taken into account~\cite{Fortov2005PR}, 
\begin{equation}
{F_{id}} = \sqrt\frac{2}{\pi}\left(\frac{T_i}{e}\right)^2 
ln\left(\frac{4M_{i}}{\eta\beta} \right)\frac{\beta^2}{M_{i} }
\end{equation}

\begin{figure}
\includegraphics[width=0.95\linewidth]{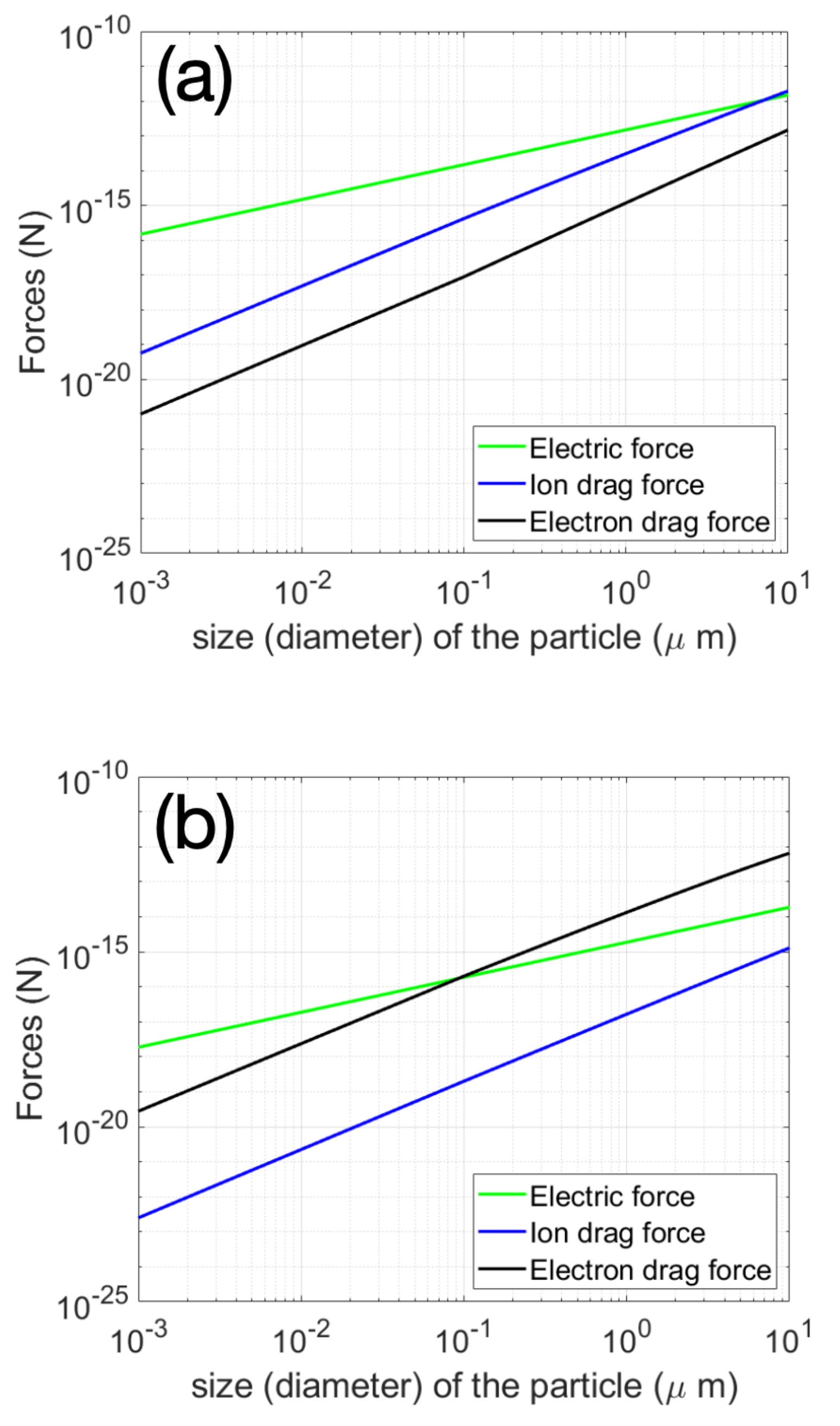}
\caption{The variations of electric force, ion drag force and electron drag force with particle size (diameter) in (a) non-thermal plasma and (b) thermal plasma. In the non-thermal plasma regime, the ion drag force always dominates over the electron drag force over the entire range of particle size under consideration. But in this regime, the electric force is the most dominant one. On the other side in thermal plasma, the electron drag force always dominates over the ion drag force over the entire range of particle size. However, in this regime, the electric force dominates over electron drag force for small particles, but there is a cross-over at $\sim$ 100 nm particle size after which electron drag force dominates over electric force. The plasma parameters used for these plots are discussed in the text.}
\label{force_estimation_plot}  
\end{figure}

\begin{figure}
\includegraphics[width=0.95\linewidth]{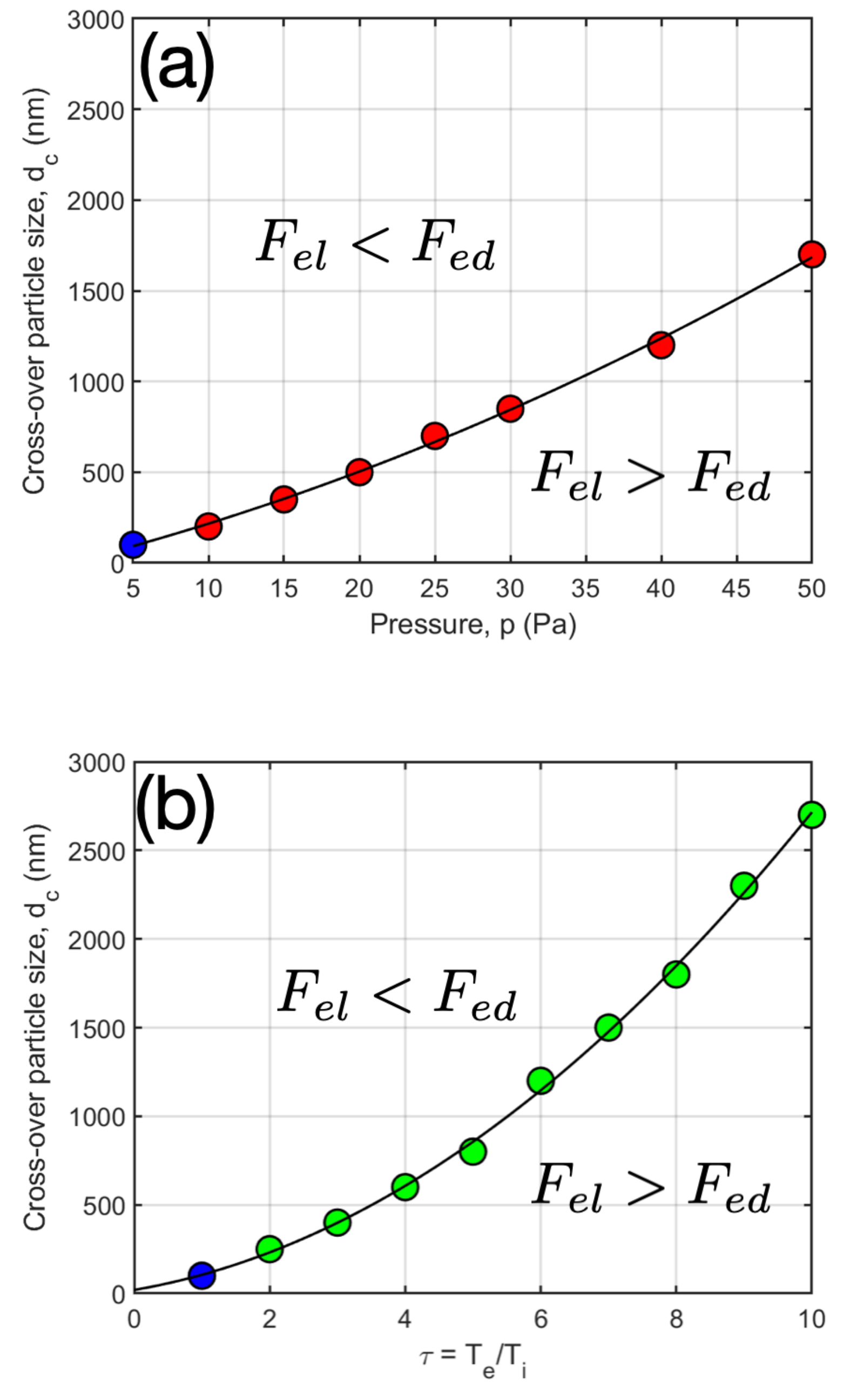}
\caption{(a) The variation of the cross-over particle size (when $F_{el} = F_{ed}$ as represented by the solid line) with pressure which can be represented by the best fit quadratic equation: $d_c = 0.27 \times p^2 + 21 \times p - 17$. Here $d_c$ is the cross-over particle size and $p$ is pressure. (b) The variation of the cross-over particle size (when $F_{el} = F_{ed}$ as represented by the solid line) with electron-to-ion temperature ratio ($\tau$) which can be represented by the best fit quadratic equation: $d_c = 20 \times \tau^2 + 65 \times \tau + 20$. The data point with blue circle in both figures represents the baseline configuration for thermal plasma ($\tau = 1$) at 5 pa gas pressure.} 
\label{force_cross_over}  
\end{figure}

Here, $M_{i} = u_i/v_{i}$ is the thermal mach number for ions and $\beta = z\tau/\xi$ is the scattering parameter. 
To highlight the new results of our work, the ion drag-to-electron drag force ratio variation over particle size is shown in Fig.~\ref{force_ratio}. It is found that electron drag force is much stronger than ion drag force in the thermal state of the EUV photon induced plasma under operational conditions. 

The variation of the electric, ion drag and electron drag forces with particle size is shown in Fig.~\ref{force_estimation_plot}. At the beginning of the pulse when the plasma is non-thermal ($\tau \sim 400$) with background gas pressure at 5 pa, the electric and ion drag forces dominates over the electron drag force by 1-2 orders of magnitude. A cross-over diameter $\sim 7 \mu$m is observed between electric and ion drag forces below which electric force dominates over ion drag force and above which opposite occurs. However, at the end of the pulse with thermal plasma condition ($\tau \sim 1$) and same background pressure at 5 pa, the electron drag force is always higher than the ion drag force by almost 3 orders of magnitude. Furthermore, the important fact is that in this weakly collisional regime, a cross-over particle size $d_c \sim 100$ nm is obtained. For smaller particle size ($d < d_c$) the electric force dominates over electron drag force and for bigger particle size ($d > d_c$) the electron drag force dominates over electric force. This is because $F_{el} \sim a$ and the electron drag force $F_{ed} \sim Z^2 \sim a^2$. Fig.~\ref{force_cross_over} shows such cross-over particle size ($d_c$) variation with pressure and electron-to-ion temperature ratio respectively. Each marker in Fig.~\ref{force_cross_over}a represents the cross-over particle size between electric force and electron drag force in the thermal plasma at the end of the pulse for a particular pressure. One such cross-over particle size (100 nm) is shown in Fig.~\ref{force_estimation_plot}b for 5 pa gas pressure. Similar cross-over size estimations are made for different pressures until 50 pa and a best fit was made as solid line as mentioned in the caption. Similar estimations have been made for varying electron-to-ion temperature ratio at a particular pressure (5 pa). From the above plots, it is possible to classify the particle sizes for which electron drag force can dominate over electric force under typical operational conditions.


In conclusion, it is found that particle transport within each EUV pulse strongly depends on the plasma regime. At the beginning of the pulse when the plasma is non-thermal, the electric and ion drag forces dominates over the electron drag force. However, at the end of the pulse when the plasma becomes thermal, the electron drag force dominates over the ion drag force for all size of particles. The electron drag force becomes most dominant (stronger than electric force) for bigger size particles. It is to be noted that this work is based on single pulse analysis. In case of multiple pulse scenario, residual charge may play an important role and its influence on the steady state charge as well as drag forces estimations should be performed self-consistently. It is also important to perform dedicated experiments for successful realization of the predictions made in this work for EUV plasma and this has been kept as future work.



$*$ Corresponding author: manis.chaudhuri@asml.com

The data that support the findings of this study are available from the corresponding author upon reasonable request.

\end{document}